\documentclass[aps,prl,reprint,superscriptaddress]{revtex4-2} 
\usepackage[pdfborder={0 0 0}, colorlinks=true, linkcolor=blue, citecolor=blue, urlcolor=blue]{hyperref}
\usepackage{bm}
\usepackage{graphicx}
\begin{document}


\title{Time-dependent DFT-based study of bacteriochlorophyll \textit{a} optical properties within the B800 part of \textit{Rhodoblastus acidophilus} light-harvesting complex}

\author{Evgenia A. Kovaleva}\email{kovaleva.evgeniya1991@mail.ru}
\affiliation{Federal Research Center Krasnoyarsk Science Center SB RAS, Krasnoyarsk, Russia}

\author{Lyudmila V. Begunovich}
\affiliation{Federal Research Center Krasnoyarsk Science Center SB RAS, Krasnoyarsk, Russia}

\author{Maxim M. Korshunov}
\affiliation{Federal Research Center Krasnoyarsk Science Center SB RAS, Krasnoyarsk, Russia}
\affiliation{Kirensky Institute of Physics, Federal Research Center Krasnoyarsk Science Center SB RAS, Krasnoyarsk, Russia}



\begin{abstract}
We use time-dependent density functional theory-based approaches, TD-DFT and TD-DFTB, to investigate the optical absorption of B800 part of \textit{Rhodoblastus acidophilus} light-harvesting\linebreak complex~2 (LH2). Both methods are shown to give qualitative agreement with experimental spectra for a single BChl\:\textit{a} molecule and for the optimized structure of B800 complex containing nine of such molecules. We proved the absence of any sizable effects originating from the interaction between adjacent molecules, thus optical features of B800 LH2 part shouldn’t be attributed to the structural organization of pigments. In addition, time-dependent procedure itself was found to be crucial for the correct description of BChl\:\textit{a} absorption spectrum.

\textit{\textbf{Keywords}} -- light-harvesting complex, purple bacteria, DFT, DFTB, TD-DFT, optical absorption spectra
\end{abstract}

\maketitle


\section{I. Introduction}

Photosynthetic processes keep the interest of researchers worldwide for decades. The shortest but still the most crucial step providing the energy for further biochemical reactions is the light absorption~\cite{Yakovlev} by structures called light-harvesting (LH) complexes. 
These structures consist of pigments bound by protein chains~\cite{Mirkovic}. 
LH2 complex of \textit{Rhodoblastus acidophilus}, also known as purple nonsulfur bacteria, is a well-known model object for study of LH absorption spectra~\cite{Maity,Saga,Swainsbury,Frigaard2006,Anda,Qian,Segatta} due to its rather simple structure: 9 rhodopin $\beta$-D-glucoside and 27 bacteriochlorophyll \textit{a} (BChl\:\textit{a}) molecules put together by 18 altering protein chains~\cite{Cherezov}. 
BChl\:\textit{a} molecules are presented as two discrete structures: a closely packed ring of eighteen BChl\:\textit{a} with their CH-planes lined up along the C\textsubscript{9} symmetry axis and a loosely packed ring of nine BChl\:\textit{a} with CH-planes oriented perpendicular to the C\textsubscript{9} axis. Such rings are responsible for characteristic spectral features of LH2 giving near infrared absorption peaks rising from Q\textsubscript{y} absorption peaks of BChl\:\textit{a} pigments at 850 (B850) and 800 nm (B800), respectively. In our work we attempt to develop a simplified model as an alternative to the widely used QM/MM approach in order to reproduce optical properties of LH2 complex while being computationally effective. We start from the description of B800 ring separately as it is less complicated and may be useful for creating a model for further studies of other parts of LH complexes.

Light-harvesting complexes are known to be tricky to describe, especially due to the pigment molecules as the interaction between them usually leads to the emergence of excitons characteristic for the system as a\linebreak whole~\cite{Mirkovic,Curutchet,Gall,Renger,KellChen,Linnanto,Strumpfer,Cupellini2023,Bose}. 
This effect was previously shown for B850 ring of LH2~\cite{Segatta,Curutchet,Cupellini2023,Bose}. 
In our recent work~\cite{Begunovich2024} we ignored possible excitonic effects since the distance between two BChl\:\textit{a} molecules in B800 ring is too large to give any collective excitations in agreement with~\cite{Curutchet,Fujimoto} so the change in Q \textsubscript{y} peak position of B800 with respect to that of BChl\:\textit{a} molecule~\cite{Frigaard1996} should be attributed to the Mg-porphyrin plane distortion caused by the interaction with protein residues. Here we use time-dependent (TD) calculations within the density functional theory (DFT) framework to prove the conjecture. 

\section{II. Computational methods and models}

We used 2FKW PDB structure~\cite{2FKW} from the study of Cherezov et al.~\cite{Cherezov} as the starting point for our calculations and took nine BChl\:\textit{a} molecules forming B800 ring without any protein residues so the impact of protein surroundings could be eliminated during the optimization procedure. Thus, we were able to examine the effect of interaction between BChl\:\textit{a} itself. Both B800 and single BChl\:\textit{a} molecule taken as the reference were fully optimized until the forces acting on atoms became less than $2*10^{-4}$ Hartree/bohr.  
We used third-order version of DFTB along with 3OB parameters~\cite{LuGaus2015,GausLu2014,Gaus2013} and Grimme D3 correction for van-der-Waals interactions~\cite{Grimme2010,Grimme2011}, as implemented in DFTB+ package.  

Since the Hartree-Fock (HF) exchange was shown to have a prominent impact on the BChl\:\textit{a} optical properties~\cite{Begunovich2024}, here we chose HSE06 hybrid functional~\cite{Krukau} for further TD-DFT calculations. 
These calculations were performed in VASP package~\cite{Kresse} using GGA-PBE~\cite{Perdew} fully optimized geometry in periodic boundary conditions (PBC) with lattice vector \textit{b} set to match the experimentally known distance of 21.1 Å between adjacent molecules~\cite{Papiz} and large vacuum intervals set in other directions in order to eliminate any artificial interactions.  
Plane wave basis set and PAW method~\cite{Blochl} were used with plane wave cutoff energy of 400 eV. 
Keeping in mind large size of the PBC unit cell, $\Gamma$-point calculations were considered to be computationally effective yet sufficient for correct description of our model.
Three different methods of absorption coefficient calculation were used according to the software used. Only TD-DFTB (time-dependent density functional functional-based tight binding) method was suitable for the whole B800 ring due to its size while isolated BChl\:\textit{a} molecule was studied by both DFTB and hybrid DFT methods.
Optical properties calculation in VASP implies the calculation of the frequency-dependent dielectric matrix~\cite{Gajdos}. 
Absorption coefficient is then calculated as

\begin{equation}\label{1}
\sigma(\omega)=\frac{\sqrt{2 \omega}}{c} \sqrt{\sqrt{\varepsilon^{\prime 2}(\omega)+\varepsilon^{\prime \prime 2}(\omega)}-\varepsilon^{\prime}(\omega)},
\end{equation}
where $\varepsilon^{\prime}(\omega)$ and $\varepsilon^{\prime \prime}(\omega)$ 
are real and imaginary parts of the dielectric function, $\omega$ is the frequency.
The approach described above is based on the ground state calculation with no excited states included. To overcome this drawback, time-dependent (TD) procedure is required. Linear-response TD method is based on solving the Casida equation,

\begin{equation}\label{2}
\boldsymbol{\Omega} \mathbf{F_{\mit{I}}}=\omega^2 \mathbf{F_{\mit{I}}},
\end{equation}
 where $\boldsymbol{\Omega}$ is the response matrix that depends on the occupied and virtual Kohn-Sham orbitals and energy difference between them ($\omega$), $\mathbf{F_{\mit{I}}}$ is the eigenvector found by solving Eq.(\ref{2}) and used for calculating the oscillator strength.

In DFTB+ package both Casida method~\cite{Niehaus} and the real time propagation of electron dynamics (ED)~\cite{Bonafe} are implemented.  
Time-dependent dipole moment $\rm{\boldsymbol{\mu}}$ dependence on the electric field $ \mathbf{E}$,

\begin{equation}\label{3}
\rm{\boldsymbol{\mu}(\omega)=\alpha \mathbf{E}(\omega)} ,
\end{equation}

gives the polarizability tensor $\alpha$, and imaginary part of its trace is proportional to the absorption coefficient that we are looking for:

\begin{equation}\label{4}
\rm{\sigma(\omega)=\frac{4 \pi \omega}{c} {Im}\left[\frac{1}{3} {Tr}(\alpha)\right]} .
\end{equation}

Hereafter we denote optical calculations via the Casida equation as TD, no matter based on hybrid DFT or DFTB electronic structure, while real-time propagation of electron dynamics is denoted as ED. HSE06 notation corresponds to the absorption coefficient obtained from dielectric matrix.

\section{III. Results and discussion}

\begin{figure}[h]
 \includegraphics
{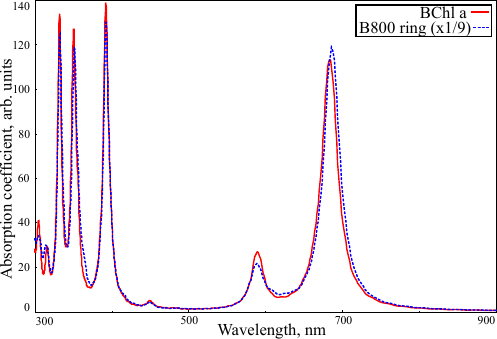}
 \caption{Absorption spectra for BChl\:\textit{a} molecule and B800 ring from ED-DFTB calculations
\label{Figure1}}
 \end{figure}

Our previous calculations of BChl\:\textit{a} spectrum~\cite{Begunovich2024} showed hybrid DFT and parametrized DFTB3 method to give reasonable results for its optical absorption spectrum: Soret low-wavelength band around 350 nm and characteristic Q region peaks are both present. However, the DFTB3 method is preferrable for large complexes of molecules in terms of future investigation of more complicated LH2 models due to its computational efficiency. Absorption spectrum of B800 ring made of nine BChl\:\textit{a} molecules without any amino acids involved is virtually identical to that of an isolated molecule and demonstrates Qy peak to be at about 700 nm compared to experimentally known position of 800 nm~\cite{Georgakopoulou}, see\linebreak
Figure~\ref{Figure1}. 

\begin{center}
 \begin{figure}[t]
 \includegraphics[width=\textwidth]
{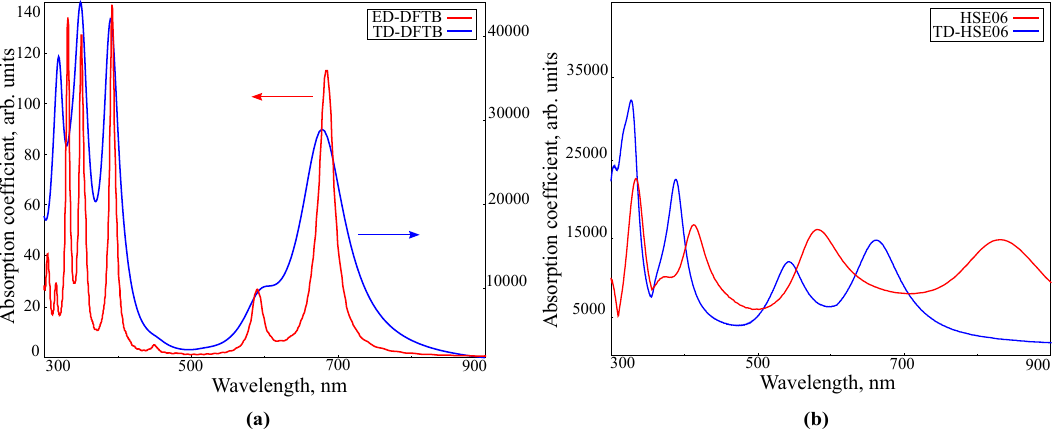}
 \caption{Absorption spectra for BChl\:\textit{a} molecule calculated using either DFTB \textbf{(a)} or HSE06 \textbf{(b)} method for electronic structure calculation \label{Figure2}}
 \end{figure}
\end{center}

This confirms our suggestion of the absence of any sizable interaction between molecules in B800 ring due to the large distance between them. To prove this, we then used the Casida equation to obtain the energies and intensities of optical transitions from DFTB and HSE06 calculations, see Figure~\ref{Figure2}.
Optical absorption spectra obtained from TD-DFTB calculations reproduce data from electron dynamics calculations and available literature~\cite{Vegte,Cupellini2018}. 
This result should be mainly attributed to the well-tested interatomic parameters designed to represent the properties of biological Mg-porphyrin compounds like chlorophyll. Only the lowest energy transition possible was calculated for B800 due to the computational expenses as there are a lot of forbidden transitions with zero intensity. The transition corresponds to Q\textsubscript{y} peak in absorption spectrum and matches other data well. Bathochromic shift of BChl\:\textit{a} Q\textsubscript{y} peak in the absorption spectrum of B800 part of LH2 complex should then be fully attributed to the structural distortions rising from van-der-Waals interaction with protein chains.

\onecolumngrid
\twocolumngrid

\begin{figure}[h]
 \includegraphics[width=0.48\textwidth]
{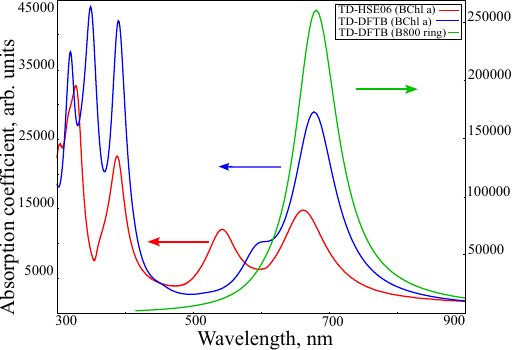}
 \caption{Absorption spectra for BChl\:\textit{a} molecule and B800 ring obtained from DFTB3 and HSE06 calculations by solving the Casida equation
\label{Figure3}}
 \end{figure}

TD-HSE06 calculation, however, demonstrates hypsochromic shift of Q\textsubscript{x} and Q\textsubscript{y} peaks with respect to ones from the frequency-dependent dielectric matrix so we can’t avoid using time-dependent procedure while investigating optical properties of pigments like BChl\:\textit{a}.\linebreak
Figure~\ref{Figure3}
demonstrates absorption spectra calculated with both methods for BChl\:\textit{a} molecule as well as DFTB results for B800 ring. Position of Q\textsubscript{y} peak is of particular interest as it’s responsible for the characteristic features of LH2 spectrum as a whole. Q\textsubscript{y} peak is located at $\sim$660-680 nm in all time-dependent calculations while TD-HSE06 gives higher energy for Q\textsubscript{x} peak, see
Table~\ref{Table1}.

\begin{table}
\caption{Absorption maxima in Q region for BChl\:\textit{a} molecule (in brackets, for the B800 ring) calculated via\linebreak different methods\label{Table1}}
\begin{ruledtabular}
\begin{tabular}{ lcc }
\textbf{Level of}& \textbf{Location of Q\textsubscript{x}}&\textbf{Location of Q\textsubscript{y}}\\
\textbf{theory}&\textbf{maximum, nm}&\textbf{maximum, nm}\\ \hline
\textbf{ED-DFTB}&  590 (589)& 683 (686)\\ \hline
\textbf{TD-DFTB}& 596& 677 (681)\\ \hline
 \textbf{HSE06}&  581& 831\\ \hline
\textbf{TD-HSE06}& 543& 661\\
\end{tabular}
\end{ruledtabular}
\end{table}

\section{IV. Conclusion}

We calculated the absorption spectra of BChl\:\textit{a} molecule as the part of B800 ring of the LH2 complex of \textit{Rhodoblastus acidophilus} within time-dependent\linebreak DFT-based methods: DFTB3 and HSE06. Time-dependent calculations confirm that (i)~even though there are no collective excitations in B800 ring, the absence of any sizable effects originating from the interaction between adjacent molecules confirms the suggestion that optical features of B800 LH2 part are induced by the protein surroundings, not by the structural organization of pigments.\linebreak (ii)~Time-dependent procedures, however, are crucial for the correct description of BChl\:\textit{a} molecule itself as ground state calculations underestimate the position of BChl\:\textit{a} Qy peak by 170 nm. (iii)~DFTB method proved to give reasonable results comparable to hybrid DFT so one can safely use it for larger models containing Mg-porphyrin compounds.


\section{Authors contribution}
E.A. Kovaleva – calculations, data processing, discussion, writing; L.V. Begunovich – resources, calculations, discussion; M.M. Korshunov – conceptualization, discussion, writing.

\section{Funding}
This work was supported by the state assignment of the Ministry of Science and Higher Education of the Russian Federation.

\section{Conflict of interest}
Authors declare no potential conflict of interest.

\section{Acknowledgments}
\begin{acknowledgments}
We thank V.F. Shabanov for useful discussions. \linebreak  Authors would like to thank Information Technology Centre, Novosibirsk State University for providing access to their supercomputers. L.V.B.  would like to thank Irkutsk Supercomputer Center of SB RAS for providing the access to HPC-cluster ``Akademik V.M. Matrosov'' (Irkutsk Supercomputer Center of SB RAS, Irkutsk: ISDCT SB RAS; http://hpc.icc.ru)
\end{acknowledgments}

\end{document}